\documentstyle[preprint,aps]{revtex}

\tightenlines
\input{psbox.tex}

\begin{document}

%
\title{Vacancy-assisted domain-growth in asymmetric binary alloys: a Monte Carlo 
study}
\author{Marcel Porta, Eduard Vives and Teresa Cast\'an}
\address{
Departament d'Estructura i Constituents de la Mat\`{e}ria,
Facultat de F\'{\i}sica, \\Universitat de Barcelona, Diagonal 647,
E-08028 Barcelona, Catalonia.}
\date{\today}
\maketitle 

\begin{abstract} 

A Monte Carlo simulation study of the vacancy-assisted domain-growth in 
asymmetric  binary alloys is presented.  The system is modeled
using a three-state ABV Hamiltonian which includes an asymmetry term, not
considered in previous works.  Our simulated system is a stoichiometric
two-dimensional binary alloy with a single vacancy which evolves according to
the vacancy-atom exchange mechanism.  We obtain that, compared to the 
symmetric case, the ordering process slows down dramatically.  Concerning 
the asymptotic
behavior it is algebraic and characterized by the Allen-Cahn growth 
exponent $x=1/2$.  The late stages
of the evolution are preceded by a transient regime strongly affected by both
the temperature and the degree of asymmetry of the alloy.  The results are
discussed and compared to those obtained for the symmetric case.

\end{abstract} 


\vspace{0.5cm}

\pacs{PACS numbers 64.60.Cn, 61.70.Bv, 81.30.Hd}


\section{Introduction}

The study of domain-growth is a prototype problem in out-of-equilibrium
statistical mechanics.  Besides its fundamental interest, it has many
technological implications in areas like metallurgy, semiconductors, surface
physics, ...etc.  Despite during the last 15-20 years domain-growth has been the
subject of a deep and continuous investigation \cite{Gunton83,Komura88,Bray94},
the models commonly used are still far from real materials.  A typical
experimental situation occurs when a binary alloy undergoing an order-disorder
transition is rapidly cooled from the high-temperature disordered-phase to the
low-temperature ordered-phase.  The growth of the ordered domains, subsequent to
the quench, is a complex phenomenon in which different factors may play an
important role:  details of the atomic diffusion, topological defects,
temperature, degeneration of the ordered phase, interface structure, impurities,
etc...  It is important to realize that the information relative to the
influence of each single factor is not always available from the experiments and
therefore alternative reliable methods such as numerical simulations may be very
useful.

Many Monte Carlo simulation studies based on the Ising-like lattice models
assume a very simplified picture of the alloy:  i.e., perfect, without defects
and with an ordering mechanism based on the neighboring atom-atom exchange.
The exchanges proposed are accepted or rejected according to the Metropolis
algorithm.  The results \cite{Phani80,Sahni81,Fogedby88} obtained are in
agreement with the phenomenological Allen-Cahn curvature-driven theory
\cite{Allen79} for the time evolution of the mean-size of the ordered domains,
i.e.  $R(t) \sim t^x$, with $x=1/2$.  Moreover, evidence has accumulated that
the ordering process exhibits dynamical scaling and that the late-time growth
behavior is subjected to a high degree of universality \cite{Mouritsen90}.
Concerning the experiments, only a small number have been designed with the aim
to check the validity of the growth-law and one may say that, in general, they
are compatible with the growth exponent $x=1/2$, at least, for quenching
temperatures below but close to the equilibrium ordering temperature.  In
reference \cite{Frontera94} the authors report a very complete summary of the
experimental data available in the literature.

Concerning binary alloys, atomic diffusion is the underlying microscopic
mechanism inherent to any ordering process.  Moreover, it is accepted that the
atomic diffusion occurs via vacancies.  In this sense, quite recently, the
possibility for the more realistic vacancy-assisted domain-growth has been
investigated \cite{Vives92,Vives93,Frontera93}.  The model description is based
on the three-state ABV lattice model \cite{Frontera94} and the microscopic
dynamical mechanism accounts for the atom-vacancy exchanges only.  The most
important result is the prediction for growth exponents greater than the
standard Allen-Cahn value.

In this paper we make another step towards the study of real alloys.  In most
theoretical studies it has been assumed that the alloys are symmetric, i.e.  
the
Hamiltonians used are invariant under the exchange of the two species, A and B,
forming the alloy.  In the framework of pure Ising models with pair interactions
the asymmetry term appear as a field term (chemical potential) which is usually
neglected.  This turns out to be irrelevant for the models without vacancies
since the conservation law for both A and B particles makes the field term to be
constant.  Nevertheless, for models including vacancies, such asymmetry terms
should be considered.

The motivation for this study comes from the fact that in many binary alloys the
vacancies exhibit a tendency to locate preferentially in one of the possible
different equivalent sublattices (into which the ordered structure can be
decomposed) \cite{Neumann76}.  Our main goal is to modelize such behavior by
taking into account the corresponding asymmetry term in the ABV model, and
perform Monte Carlo simulations of the vacancy assisted domain growth.  Although
the present simulations are restricted to a two-dimensional lattice, we expect
that the general conclusions will apply to real alloys.

The results obtained may be summarized as follows:  The main effect of the
asymmetry parameter is to decrease dramatically the speed of the ordering
process.  The specific interaction between vacancies and interphases is
repulsive and, therefore, the vacancy migration from the ordered bulk to the
interphases, which is required for the domains to grow, needs to be thermally
activated.  The associated energy barrier gives rise to a temperature-dependent
transient previous to the true asymptotic behavior.  Moreover, the process of
ordering inside the domains lasts until long times.  Nevertheless, we obtain
that the long-time domain-growth behavior is algebraic, and compatible with the
Allen-Cahn theory with an exponent $x=1/2$ at all the temperatures and degrees
of asymmetry studied.  This result is different from previous studies of
vacancy-assisted dynamics of domain-growth in symmetric binary alloys where the
standard $x=1/2$ value is found only at temperatures $T \rightarrow T_c$ while,
at moderate and low temperatures, the exponent is definitively much larger than
$1/2$ and tends to $1$ for $T \rightarrow 0$.

The remaining of the paper is organized as follows.  Next, in section II, we
briefly summarize the ABV model Hamiltonian and discuss the behavior of the
vacancies in terms of the model parameters.  In the same section we also define
the model dynamics.  In section III we present the details of the Monte Carlo
simulations and provide the definition of the relevant quantities computed.
Results are presented and discussed in section IV.  Finally, in Section V we 
conclude.

\section{The model}
\subsection{The ABV Hamiltonian}

The ordering configurations of an AB alloy with a constant concentration of
vacancies, are described in terms of three state spin-1 variables $S_i=1,-1,0$,
defined on each lattice site $i=1,2....N$.  $S_i=1$ means that site $i$ is
occupied by an A atom, $S_i=-1$ by a B atom and $S_i=0$ by a vacancy.  We shall
restrict to 2d-square lattices with a constant number of particles ($N_A$,$N_B$)
and vacancies ($N_V$).  Considering only nearest neighbor (n.n.)  pair
interactions, the Hamiltonian for the ABV model can be written as
\cite{Frontera94,Frontera97}:

\begin{equation}
{\cal   H} = J\sum_{\langle ij \rangle } S_iS_j +    K\sum_{\langle ij 
\rangle } S_i^2 S_j^2 +
L\sum_{\langle ij \rangle}(S_i^2S_j + S_iS_j^2)
\label{hamiltonian}
\end{equation}
where the sums extend over all n.n.  pairs.  It contains three independent
parameters:  $J$, $K$, and $L$.  The parameter $J$ determines the ordering of
the system.  We take $J>0$ in order to ensure that the ground state will be
antiferromagnetically ordered and formed by two alternating sublattices
(A-sublattice and B-sublattice).  In what follows we shall take $J$ as the unit
of energy.  The parameter $L^*\equiv L/J$ accounts for the energy difference
between AA and BB bonds, i.e..  it controls the asymmetry of the alloy.  Finally,
the parameter $K^*\equiv K/J$ controls the energy of the bonds involving
vacancies, and therefore the vacancy-vacancy specific interaction.  In Table
\ref{TABLE1} we have summarized the different bond energies and its
corresponding excess-energy with respect to the ordered AB bond.

The  behavior of the  vacancies  during the  ordering  process  depends on both
parameters   $K^*$  and  $L^*$,  and  may  be  understood   from  the  following
considerations:

\begin{enumerate}

\item For given values of $N_A$, $N_B (=N_A)$ and $N_V$, the asymmetry parameter
$L^*$ controls the tendency for the vacancies to locate preferentially in one of
the two sublattices.  If, for instance, $L^*>1$ ($L>J$) the energy of the BB
bond is lower than the energy of the AB bond.  The vacancies, in particular
those initially located in the A-sublattice, will show a tendency to migrate to
the B-sublattice.  The same argument holds if $L^*<-1$ for the A-sublattice.

\item  Furthermore  for $|L^*|>1$, the tendency for the vacancies to concentrate
in a given sublattice, determines the specific interaction between vacancies and
antiphase  boundaries  (APB). The APB's are interphases separating 
thermodynamically equivalent 
ordered domains with the same absolute value of the corresponding order 
parameter.  These  APB's are always present during the
domain-growth  process  and are  formed by a  sequence  of AA and BB bonds.  For
values of $|L^*|<1$, the AA and BB bonds are unfavorable with respect to the AB
bond, and therefore the vacancies tend to concentrate at the APB's.  But if, for
instance,  $L^*>1$  the BB bond is  energetically  the most  favorable  and the
vacancies, located preferentially in the B-sublattice, will not show any natural
tendency to go to the APB's (notice that the AA unfavorable  bonds in the APB's
can only be broken by vacancies located in the A sublattice).  In short, for the
asymmetric  case  ($|L^*|>1$),  the vacancies will prefer to stay in the ordered
bulk rather than to concentrate at the APB's, and the corresponding energy
difference is proportional to $(|L^*| -1)$.

\item Finally, we notice that the behavior described above may be modified by
the specific interaction between vacancies, controlled by the parameter $K^*$.  A 
simple analysis shows that this interaction
is attractive for $K^*<1$ and repulsive for $K^*>1$.
Recently, the effect of $K^*$, in the simplest case $L^*=0$, has been
extensively studied \cite{Marcel97,Srolovitz87,Mouritsen89,Shah90}.  Here we
shall concentrate on the effect of $L^*$ only and leave for a further
investigation the interplay between $K^*$ and $L^*$ that may be very subtle.

\end{enumerate}

\begin{figure} 
\psboxto(0.8\textwidth;0cm){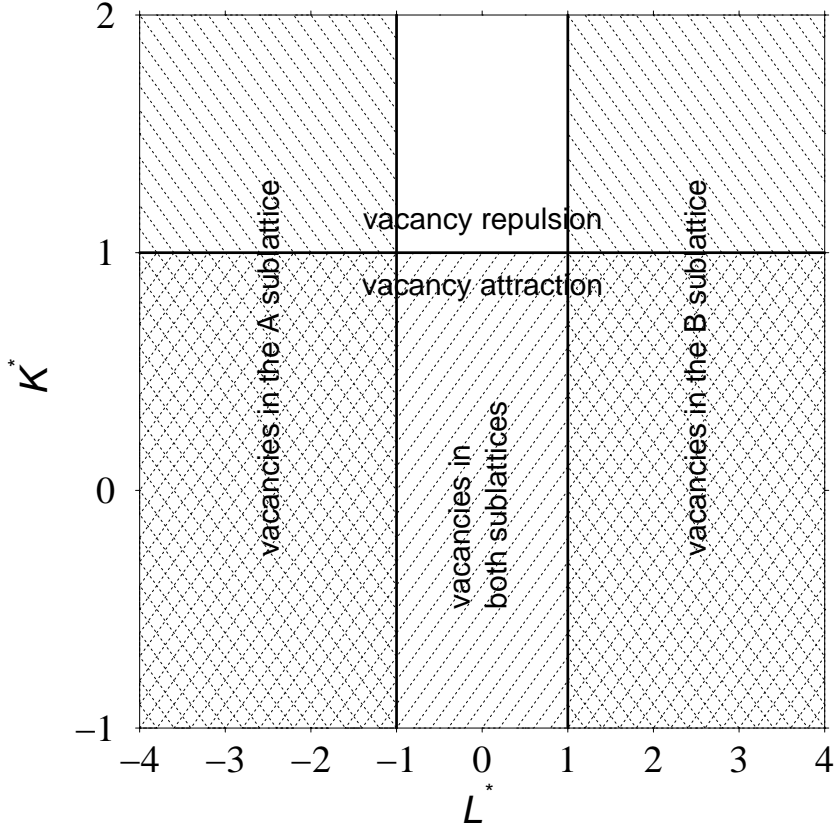}
\caption{Space of  the parameters $K^*$   and $L^*$ divided  into  the
regions where different dynamics are expected.  The solid line $K^*=1$
separates  the region of vacancy attraction  $(K^*<1)$ from the region
of vacancy repulsion $(K^*>1)$.  The solid lines $L^*=\pm 1$ separates
the region where the vacancies  tend to  precipitate at the  antiphase
boundaries $(|L^*|<1)$ from  the region of  vacancy-antiphase boundary
repulsion $(|L^*|>1)$.}
\label{FIG1}
\end{figure}

From the above considerations one finds six different regions in the space of
model parameters, as it is illustrated in Fig.\ref{FIG1}.  We notice that the
different behaviors of the vacancies are separated by the lines $|L^*|=1$ and
$K^*=1$.

\subsection{Model dynamics} 

The non-equilibrium properties of model (\ref{hamiltonian}) may be studied using
Monte Carlo computer simulation techniques \cite{Binder88}.  This requires the
implementation of a given microscopic ordering mechanism, compatible with the
conservation laws in effect.  In our case this means to preserve the number of
particles of each kind as well as the number of vacancies.  We have chosen the
vacancy mechanism:  a single vacancy is considered and only atom-vacancy
exchanges are allowed.  We call this a vacancy jump.  Proceeding as it is usual,
the vacancy jumps are proposed to n.n.  and next nearest neighbors (n.n.n)
sites, with equal probability.  This is very convenient when performing
Monte Carlo simulations in order to prevent trapping phenomena \cite{Fogedby88,Vives92}.
The attempts are accepted or rejected following the Metropolis algorithm which
only takes into account the energy difference between the initial and final
states in a given vacancy jump, that is:  $P(\Delta \cal H)$ = $\min \lbrace 1,
\exp(\frac{-\Delta \cal H}{ k_B T}) \rbrace $.  The unit of time, the Monte
Carlo step (MCS), is defined as $N$ attempts of single vacancy jumps.

In the present study the word ``barrier'' is used to denote the extra energy
needed for the vacancy to migrate from the bulk to the antiphase boundaries.
This barrier is intrinsic to model (\ref{hamiltonian}) and should not be
confused with eventual energy barriers associated to the details of the vacancy
path in a given jump between two neighboring sites.  There exist in the
literature Monte Carlo algorithms which consider the existence of these last
barriers \cite{Fultz87}.  Nevertheless, for the sake of clarity, in this first
study of domain growth in asymmetric binary alloys, we will not consider them.

\section{Simulation details}

Our simulated system is a (nearly) stoichiometric AB alloy containing a single
vacancy.  Notice that, in this case ($N_V=1$, $N_B=N_A -1$), the parameter $K^*$
is irrelevant and therefore may be neglected in model (\ref{hamiltonian}).  The
particles ($N-1$) are sitting on a square lattice of size $\ell \times \ell=N$,
subjected to periodic boundary conditions.  The main results presented
correspond to $L^*=2$ and $\ell = 600$ ($N=3.6 \cdot 10^5$).  Some additional
results for other values of the asymmetry parameter ($L^*=0$, $L^*=1$ and
$L^*=3$) are also shown.  Starting from a completely disordered configuration,
the ordering process is studied at different quenching temperatures ($T^* \equiv
k_B T / J = 1.0$, $1.25$ and $1.5$) below the critical ordering temperature
$T_c^* \simeq 2.27$.  The simulations are extended up to $\sim 4 \cdot 10^4$ MCS
and the results are averaged over $\sim 250$ independent runs.

From the information given above it follows that the concentration of vacancies
in our simulated system is $\sim 10^{-6}$. This is reasonable since typical 
concentrations of vacancies may range from $\sim 10^{-3}$ to $\sim 10^{-10}$,
strongly depending on temperature \cite{cvalloys}.  For the 
case with $L^*=0$ \cite{Frontera94}, it has been shown that the results
obtained in vacancy-assisted dynamics do not change if one increases the number
of vacancies (and the system size) maintaining its concentration constant.  This
is because of the small vacancy concentration we are dealing with which makes
the vacancy-vacancy interaction term in eq.(\ref{hamiltonian}) to be negligible
in front of the other contributions.  We expect that the same will apply in the
present non-symmetric study and, therefore, focus on the
simplest case $N_V=1$.

From the simulations we have measured the following quantities:

\begin{enumerate}

\item{\bf Long range order parameter  $m$}.  It is defined as the absolute 
value of the antiferromagnetic order parameter on the system.

\begin{equation}
m = \frac{1}{N} \sum_{i=1}^N  S_i  sign(i)
\end{equation}

where the function $sign(i)$ takes alternating $\pm 1$ values on the lattice 
in a chessboard way.

\end{enumerate}

In order to obtain  information  concerning  the path of the vacancy  during 
the
ordering  process  it is  convenient  to define the two  following  local  
order
parameters:

\begin{enumerate}

\setcounter{enumi}{1} \item{\bf Local order parameter around the vacancy 
$m_v$}.
It is defined as the absolute value of the antiferromagnetic order parameter in
a travelling ($5\times 5$) cell centered at the vacancy position.  In order to 
reduce the
numerical uncertainties partial averages over consecutive MCS have been
performed.  In the case of $L^*=2$ each value is the average over 10
(consecutive) MCS while for $L^*=3$ the same has been done over $100$ MCS.
Furthermore, averages over many (30-1000) independent runs have been performed
in order to minimize the dependence with the initial conditions.

\item{\bf Local order parameter $m_{(5 \times 5)}$}.  It is defined as the
absolute value of the antiferromagnetic order parameter in a $(5\times 5)$ cell.
The computation is done by dividing the original lattice into
cells of size $(5\times 5)$ and then averaging over all of them.

\end{enumerate}

Domain-growth can be monitored by measuring other quantities.  We shall study
the excess-energy and the width of the structure factor.

\begin{enumerate}

\setcounter{enumi}{3}

\item{\bf Excess internal energy $\Delta E(t)$}. It is defined as the excess 
internal energy per particle:

\begin{equation}
\Delta E(t) = E(t) - < {\cal H}>_T
\end{equation}

where $E(t)$ is the energy of the system (given by eq.  (\ref{hamiltonian})) at
time $t$ after the quench and $< {\cal H}>_T$ is the equilibrium energy of the
system at the temperature $T$.  At each temperature, this equilibrium energy has
been obtained on a system of $\ell=200$, starting from a completely ordered
configuration and then using the standard atom-atom exchange mechanism.  The
final value of $<{\cal H}>_T$ is the average over the last $1.8 \cdot 10^4$ MCS
of a single evolution of $2 \cdot 10^4$ MCS long.

\item{\bf  Structure  factor.}  The  structure  factor is defined as the 
Fourier transform of the correlation function.  It is written as:

\begin{equation}
S(\vec{k}) = \frac{1}{N} \left| \sum_{i=1}^N S_i {\rm exp} \left\{{\rm
i}\frac{2\pi}{a}\vec{k}\cdot\vec{r_i}\right\} \right|^2
\end{equation}

where $\vec{k}$ is a reciprocal  space vector, $a$ is the lattice spacing, 
and
$\vec{r_i}$  is the position  vector of site $i$.  We have computed the 
profiles
along the (10) and (11) (and equivalent)  directions, around the  
superstructure
peak at $\vec{k}=(\frac{1}{2}\frac{1}{2})$.

\item{\bf Structure factor width $\sigma(t)$}.  It is defined as the square
 root of the second moment of the structure factor:

\begin{equation}
\sigma(t)   =  \left    (  \sum_{q=0}^{q_{\rm  max}}   q^2
S(q) \left/ \,\, \sum_{q=0}^{q_{\rm max}} S(q)
\right. \right ) ^{1/2}
\end{equation}
where  $q\equiv  |\vec{k}-(\frac{1}{2} \frac{1}{2})|$  is  the  distance  to 
the
superstructure   peak.  The  sum  is  performed  over  the  $q$  values  in  
the
corresponding  direction  and extends up to $q_{max} \equiv 3\sigma(t)$. Note 
that 
$\sigma(t)$ and $q_{max}$ have been obtained 
in a self-consistent way. By this method we avoid the problems associated to
the fitting of a profile function and to the determination of the background 
which, in this problem, cannot be neglected and shows a time evolution.

\end{enumerate}

Then, provided scaling holds,

\begin{equation}
\Delta E(t) \sim m(t)^{-1} \sim \sigma(t) \sim R^{-1}(t) 
\label{energy}
\end{equation}
where $R(t)$ is the mean size of the ordered domains.
Finally, in order to reveal the existence of transient regimes during the
growth, it is convenient
to introduce a definition for the growth exponent, not affected by {\sl
a priori} assumptions:

\begin{enumerate}

\setcounter{enumi}{6}

\item{\bf Effective growth exponent  $x_e(t)$}.  It is defined  as the
logarithmic derivative of the excess-energy according to:

\begin{equation}
x_e(t) =   \frac{-d  \log (\Delta E(t))}{d  log (t)}   = \lim_{\nu   \rightarrow 
0}
\frac{ -\log (\Delta E(\nu t)/ \Delta E(t))}{\log \nu}
\end{equation}

where, in the computation, instead of the limit, we have taken  a small value of 
$\nu$ ($\nu=2$) and fitted a power-law to all intermediate data contained in the 
range $(t, \nu t)$.

\end{enumerate}

\section{Monte Carlo results and Discussion}

\begin{figure} 
\psboxto(0.8\textwidth;0cm){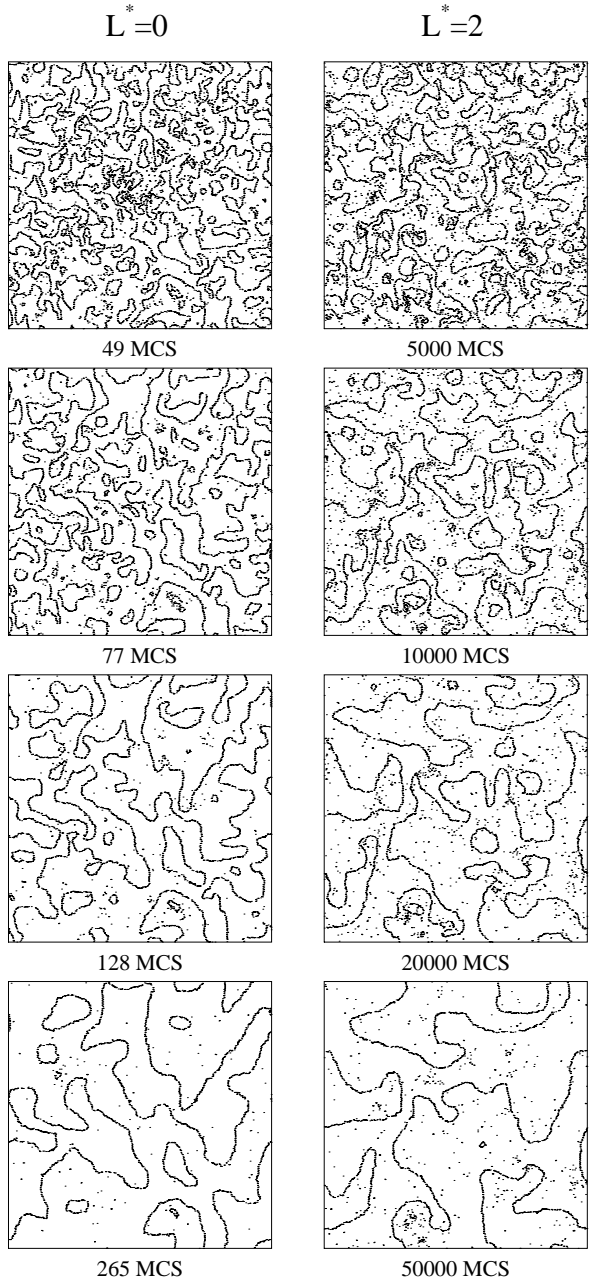}
\caption{Sequence of snapshots of the evolving domain structure for a system
with $\ell=600$, $T^*=1.0$ and two values of $L^*$. Black regions represent 
disordered sites}
\label{FIG2}
\end{figure}

In this section we present the numerical results obtained from the Monte Carlo
simulations.  Figure \ref{FIG2} shows snapshots of the configurations obtained
at $T^*=1$, for $L^*=0$ and $L^*=2$ at four different selected times.
Disordered sites are shown in black.  The snapshots for the two values of $L^*$
correspond to simulation times for which the system has the same excess internal
energy.  The overall picture of the domains is qualitatively similar.
Nevertheless, for $L^*=2$ one observes that the amount of disorder in the bulk
is larger.

\begin{figure}
\psboxto(0.8\textwidth;0cm){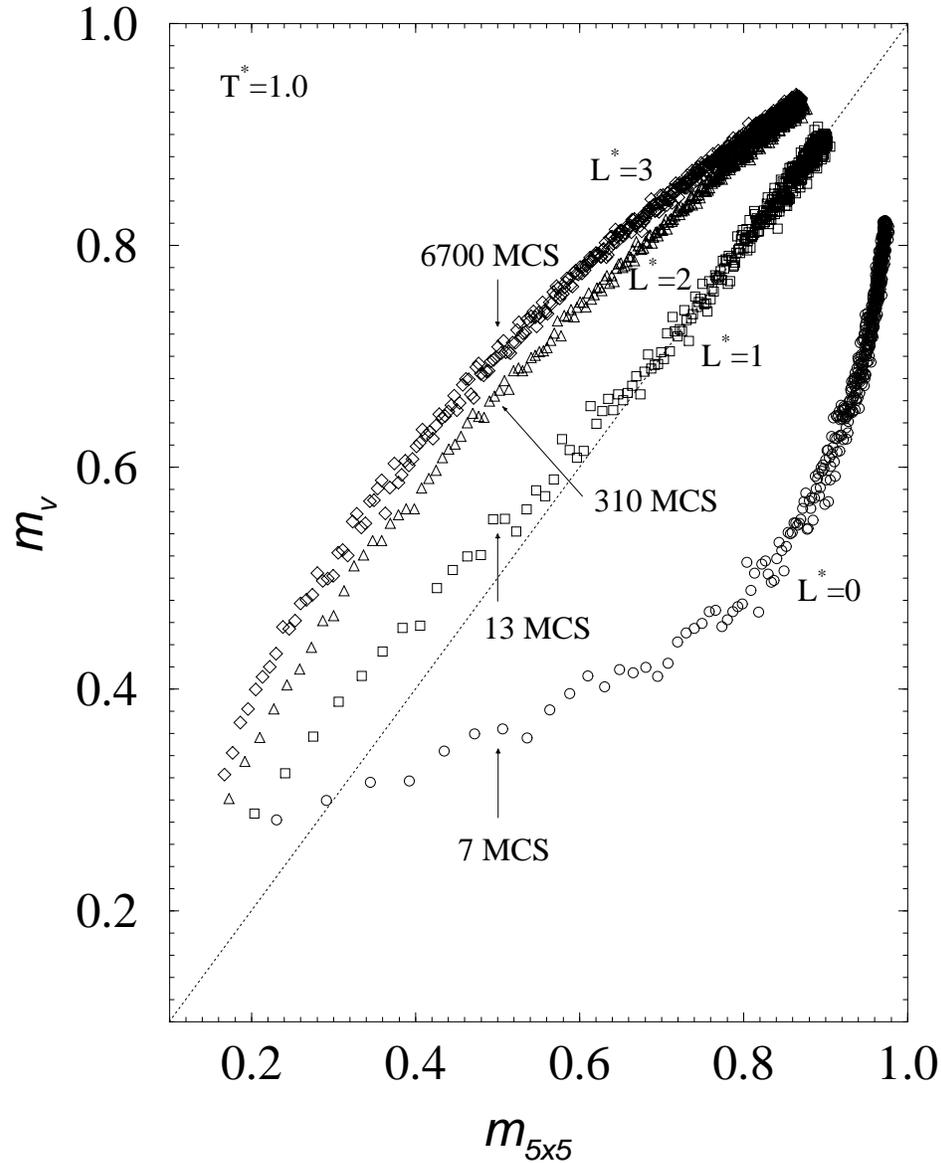}
\caption{Evolution of the local order parameter defined around the 
vacancy $m_v$
vs. the local order parameter $m_{5 \times 5}$ for a system with
$\ell=200$, $T^*=1.0$ and different values of $L^*$, as indicated. The dotted
straight line shows the behavior corresponding to a random walk of the vacancy. 
The 
arrows indicate, for each case, the number of MCS necessary to reach 
a value $m_{5 \times 5} = 0.5$.}
\label{FIG3}
\end{figure}

Prior to the study of the growth-law itself, it is interesting to analize the
behavior of the vacancy during the ordering process.  More precisely, it is
important to understand the characteristics of the path followed by the vacancy
during the vacancy-assisted ordering process which takes place as a response to
the quench, performed at $T^*$ .  This is done by monitoring the behavior of
$m_v$ vs.  $m_{(5 \times 5)}$ as it is illustrated in Fig.  \ref{FIG3} for a
system of linear size $\ell = 200$, temperature $T^*=1.0$ and four different
values of the asymmetry parameter $L^*$.  The straight line denotes a random
walk eventually followed by the vacancy in case the path is uncorrelated with
the state of order in the system.  As it can be seen in Fig.  \ref{FIG3}, for
$L^*=2$ and $L^*=3$, the order around the vacancy is clearly higher than the
average local order in the system (at all times $m_v > m_{(5 \times 5)}$).  This
is indicative of the tendency of the vacancy to stay preferentially in the
ordered regions (inside the bulk).  This behavior is opposite to the one found
for $L^*=0$ \cite{Vives92,Vives93}, where the vacancy clearly exhibits a natural
tendency to concentrate at the interphases (APB's).  Notice that, after an
initial transient, the intermediate case $L^*=1$ corresponds to the uncorrelated
random walk.  Moreover, it is clear that the process rapidly slows down as $L^*$
increases (especially for $|L^*|>1$).  This can be realized from the number of
MCS (indicated by an arrow) needed to achieve the same amount of order in the
system (for instance a value of $m_{(5 \times 5)} =0.50$).  Notice also, that in
all cases, at short times, $m_v > m_{(5 \times 5)}$.  This is because the
ordering mechanism involves the vacancy jumps exclusively and therefore, at
short times, the system can only be locally ordered around the vacancy.  The
above scenario is in full agreement with the discussion given in section II
i.e., for $|L^*|>1$ the specific interaction between the vacancy and the APB is
repulsive.  In what follows we shall mainly concentrate in the asymmetric case
with $L^*=2$.

\begin{figure}
\psboxto(0.8\textwidth;0cm){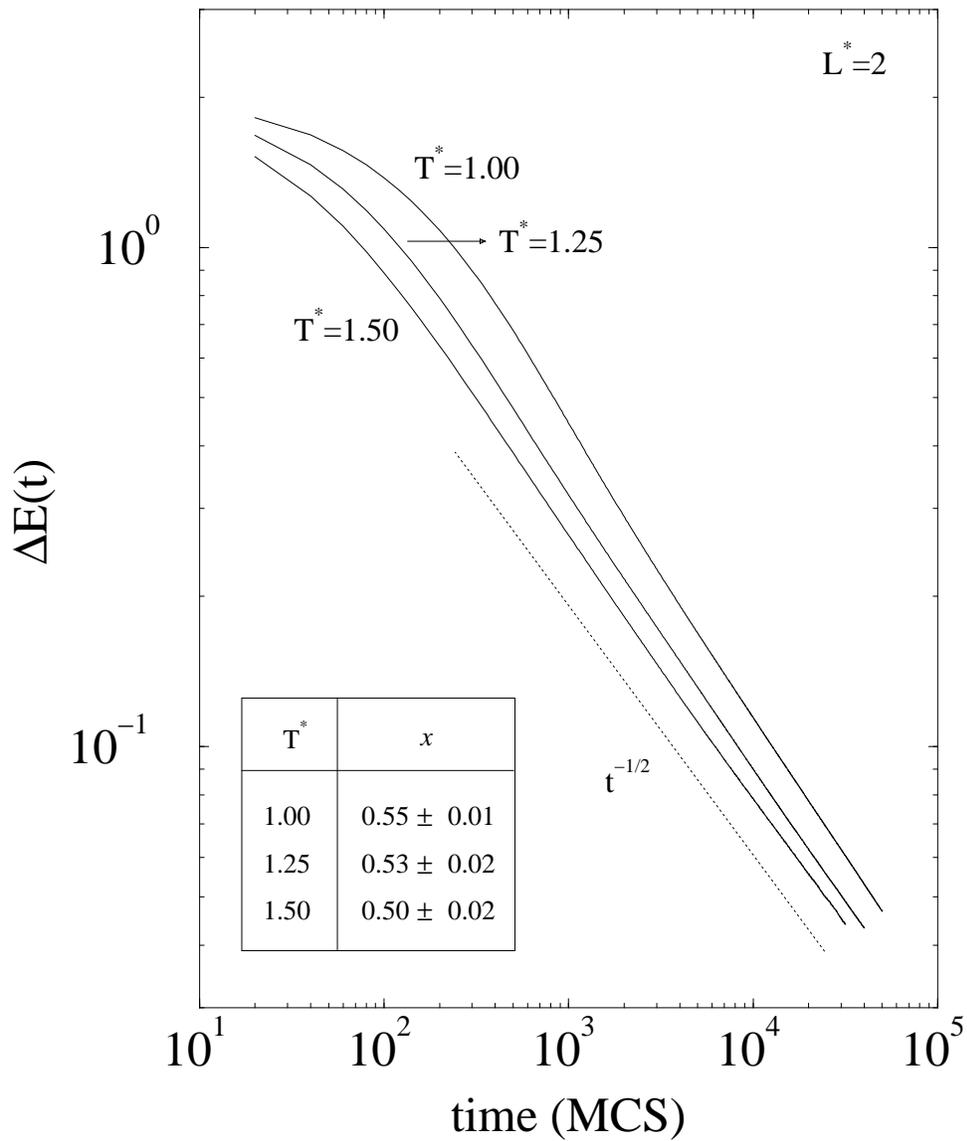}
\caption{Log-log plot of the time evolution of the excess-energy $\Delta E(t)$
for a system with $L^*=2$, $\ell=600$ and three different temperatures.
 The dotted 
straight line indicates the Allen-Cahn behavior $\Delta E \sim t^{-1/2}$. The
inset shows the fitted power-law exponents to the last 25000 MCS.}
\label{FIG4}
\end{figure}

\begin{figure}
\psboxto(0.8\textwidth;0cm){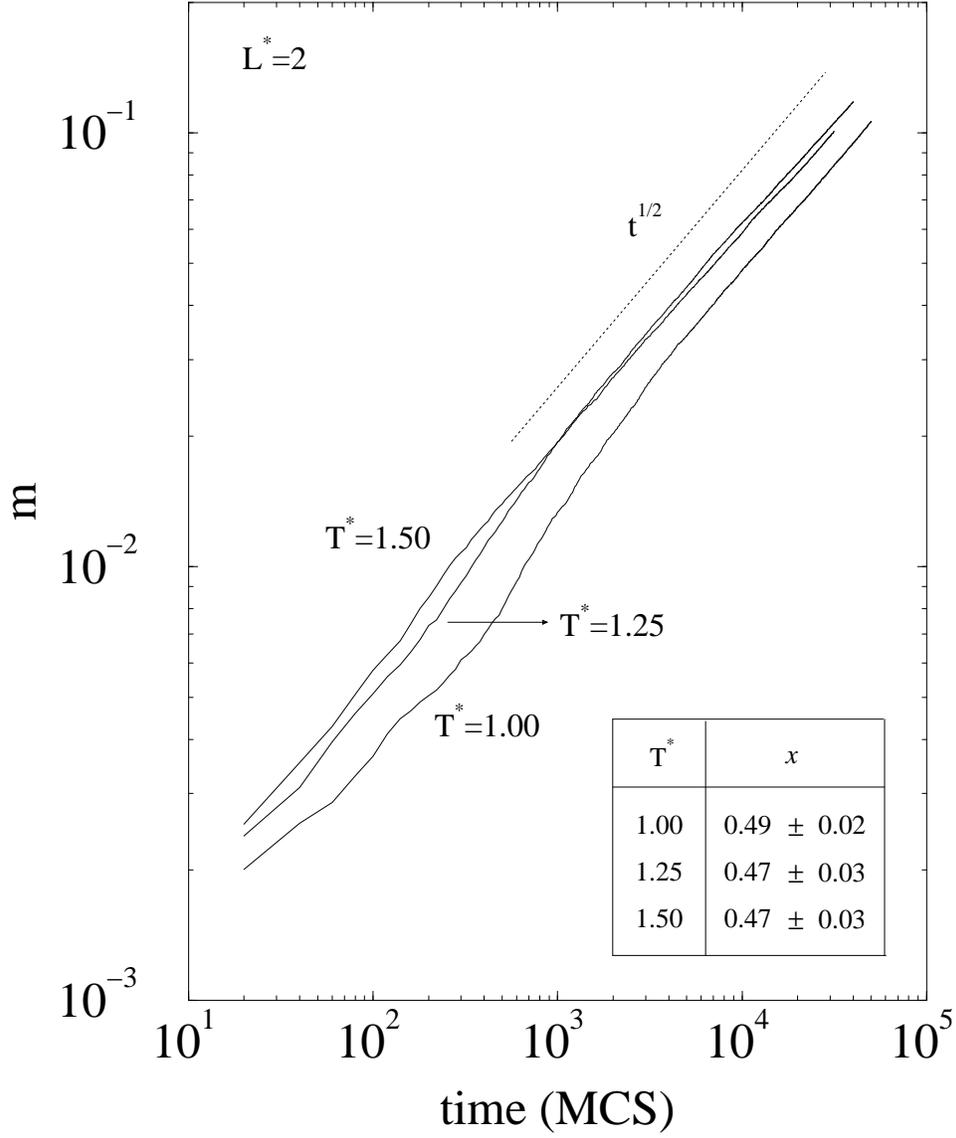}
\caption{Log-log plot of the time evolution of the long-range order parameter 
$m$, for the same temperatures than in Fig. 4.  The dotted 
straight line indicates the Allen-Cahn behavior $m \sim t^{1/2}$. The
inset shows the fitted power-law exponents to the last 25000 MCS.}
\label{FIG5}
\end{figure}

We first study the time evolution of the growing domains by following the 
behavior of $\Delta E(t)$.  This is shown, in a log-log plot, in Fig.
\ref{FIG4} for three different quenching temperatures $T^*=1.0, 1.25, 1.50$ and
$\ell =600$.  In order to verify that finite size effects do not mask the long
time behavior, we have also measured the corresponding time evolution of the
order parameter $m(t)$, which is shown in Fig.  \ref{FIG5}, for the same
three temperatures. Note that simulations have been run up to
a time ($ \sim 4 \cdot 10^4$ MCS) for which the order parameter is $m \simeq
0.1$.  This is enough to ensure that, in the long time stages studied, the
system is still in the regime of competing domains.

Results in Figures \ref{FIG4} and \ref{FIG5} indicate that the asymptotic regime
of $\Delta E(t)$ and $m(t)$ \cite{ordpar} is, for the three temperatures,
compatible with the algebraic Allen-Cahn growth-law, denoted in the figures by a
dotted straight line.  We have estimated the growth-exponents of the different
curves by least-square fitting a power-law to the last $\sim 25000$ MCS.  
They are
indicated in the insets.  All the values are compatible with the Allen-Cahn
exponent ($x=0.50 \pm 0.06$).  Nevertheless, notice that the values of the 
exponent
obtained from $\Delta E(t)$ show a slight tendency to increase as one decreases
the quenching temperature.  This has to do with the temperature-dependent
transient regime, not revealed by the behavior of $m(t)$, that may affect the
asymptotic behavior up to very long times.  Moreover, from our simulations we
have observed that, inside the domains, the equilibrium ordering is established
asymptotically and is not complete until long times (see Fig. \ref{FIG2}). 
 From 
all these
considerations it follows that, in our case, $\Delta E(t)$ has to be taken as 
an
approximate measure of the domain size evolution and that the corresponding
exponents are only orientative of the tendency of the system.  We now turn 
back 
to
the transient regime.  It is related to the existence of thermally activated
processes which hinder the growth of the ordered domains.  Indeed, once the
domain structure is formed, the domains grow only when the vacancy is located 
at
the APB's.  Since for $|L^*|>1$ the vacancy prefers to stay in the ordered bulk
(as it was discussed in section II), the migration to the APB's needs to be
thermally activated.  The corresponding energy barrier that the vacancy has to
climb can be easily evaluated and one obtains $\epsilon^* =4(|L^*|-1)=4$.  This
barrier verifies $\epsilon^* > T^*$, even for the highest temperature studied.

\begin{figure}
\psboxto(0.8\textwidth;0cm){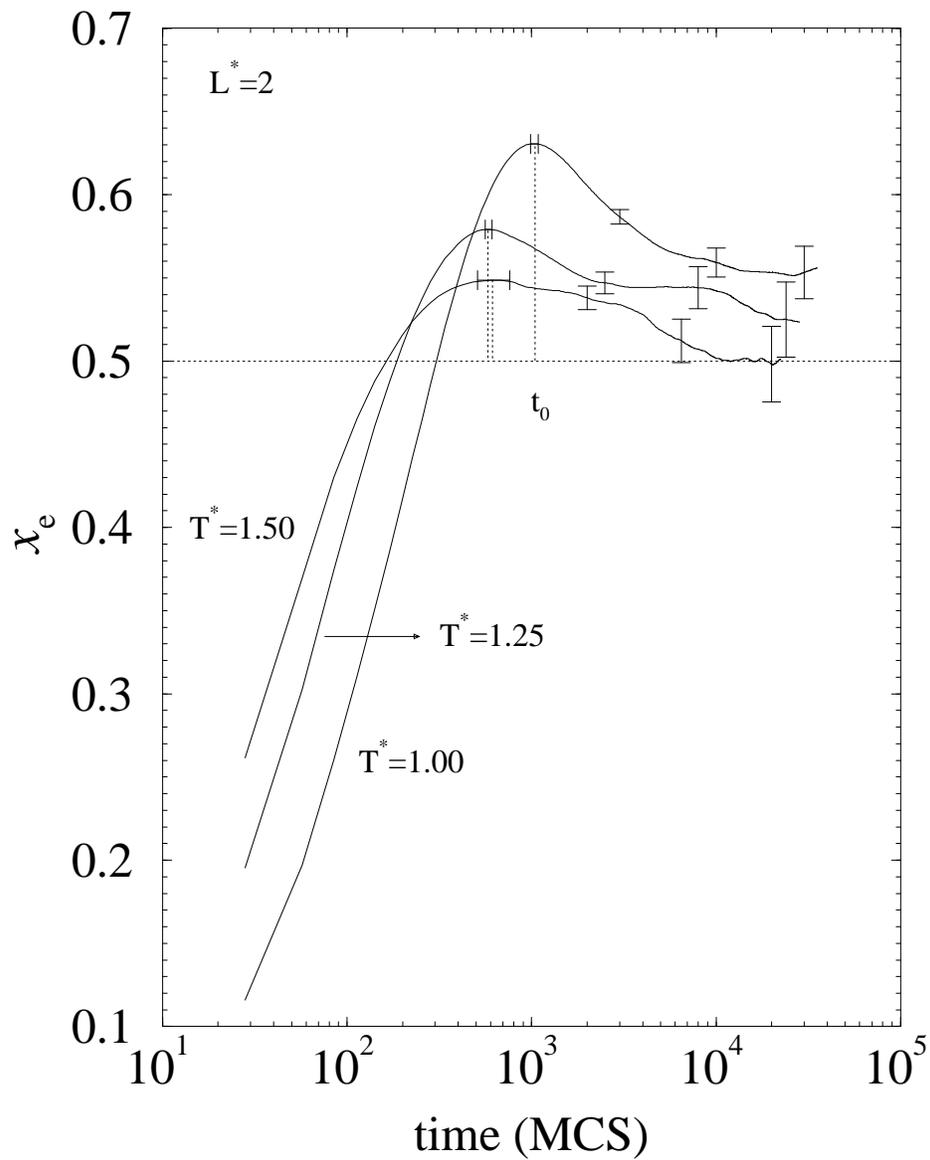}
\caption{Effective growth exponent as a function of time obtained from
the logarithmic derivative of the behavior of $\Delta E(t)$, corresponding
to the same cases as in Fig. 4. Estimations of $t_0$ and typical error bars 
are shown.}
\label{FIG6}
\end{figure}

\begin{figure}
\psboxto(0.8\textwidth;0cm){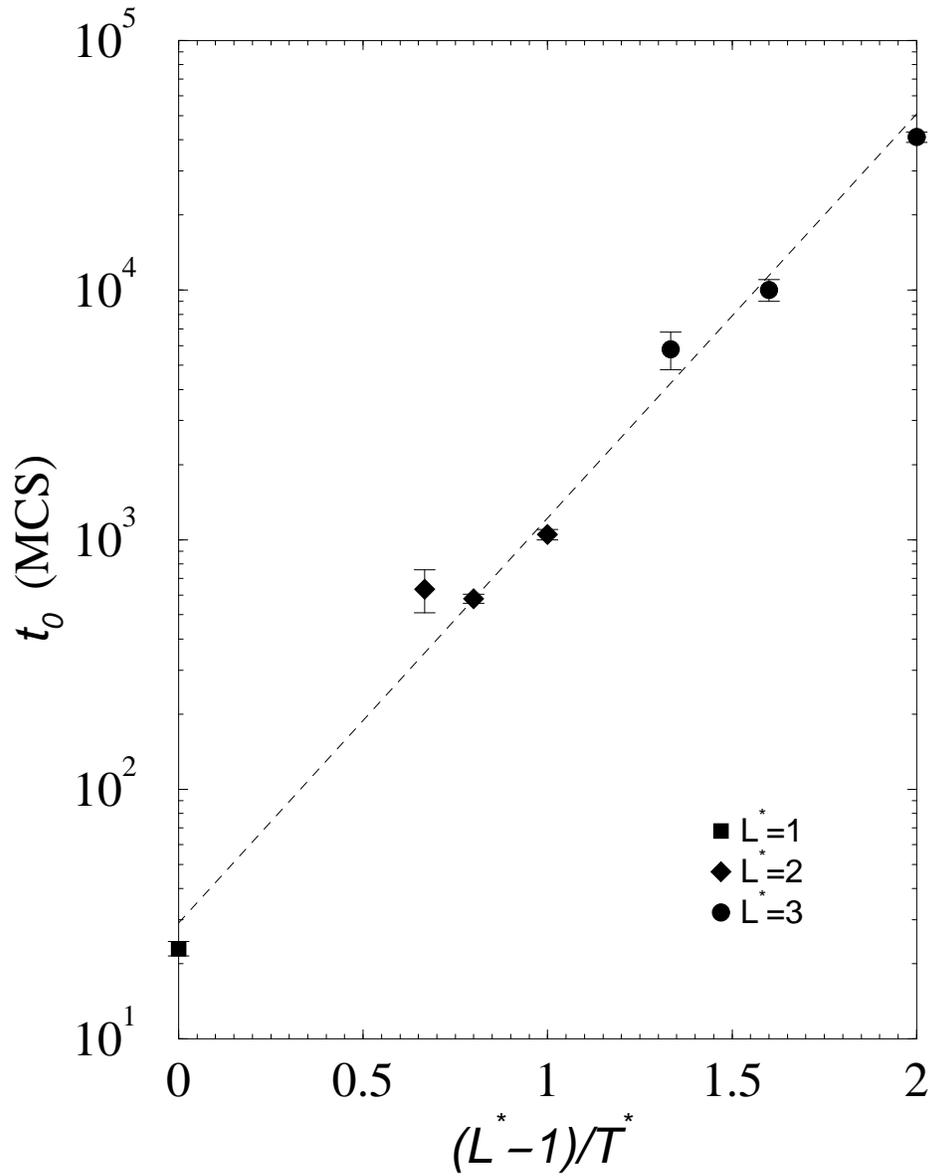}
\caption{$t_0$ as a function of $(L^*-1)/T^*$. The dashed line shows 
the fitted exponential behavior with a slope $\epsilon^*/(L^*-1) = 3.7$.}
\label{FIG7}
\end{figure}

A more quantitative analysis of the whole evolution can be extracted from the
computation of the effective exponent obtained from the logarithmic derivative
of $\Delta E(t)$, as it was defined in the previous section.  Results are shown
in Fig.  \ref{FIG6}.  The general trends are, for the three temperatures
studied, the same.  The effective exponent first increases from very small
values (at short times) up to a maximum (over $x=1/2$) at a time $t_0$ and next
decreases to reach, asymptotically the Allen-Cahn value.  On the other hand, the
position of the maximum ($t_0$) and the value of the maximum depend on
temperature:  both are larger as lower the temperature.  We have evaluated the
value of $t_0$ for the different temperatures and get $t_0= 1\cdot 10^3,
0.59\cdot 10^3$ and $0.64\cdot 10^3$ MCS for $T^*=1.0,1.25$ and $1.50$
respectively.  Moreover, the value of $t_0$ depends on $L^*$.  In fact, we have
performed simulations for $L^*=3$ (not shown in the paper) at the same 
temperatures,
and get $t_0=4.1\cdot 10^4, 1\cdot 10^4$ and $0.58\cdot 10^4$ MCS respectively.
The value of $t_0$ is, for $L^*=3$ much larger than for $L^*=2$, providing an
idea about the increasing times needed to reach the asymptotic regime when
increasing $L^*$. We expect that $t_0$ is proportional to the passing time of 
the barrier,
\begin{equation}
\label{passing}
 t_0 \sim e^{\epsilon^*/T^*} 
\end{equation}
We can check that $\epsilon^* \propto L^*-1$ by representing $t_0$ 
{\it vs.} the factor $(L^* -1) / T^*$ in a linear-log plot.  This is shown
in Fig. \ref{FIG7} for different values of  $L^*$.  The 
square corresponds to $L^*=1$, the diamonds to $L^*=2$ and the dots to
$L^*=3$. The dashed line shows an exponential fit, which renders $\epsilon^* = 
3.7 (L^*-1)$, in agreement with  the theoretical estimation given previously,
 $\epsilon^* = 4(|L^*| -1)$. The deviation of the points corresponding to the 
highest temperature ($T^*=1.5$) may be attributed to the existence of higher 
order corrections in Eqn. (\ref{passing}) due to entropic effects.

\begin{figure}
\psboxto(0.8\textwidth;0cm){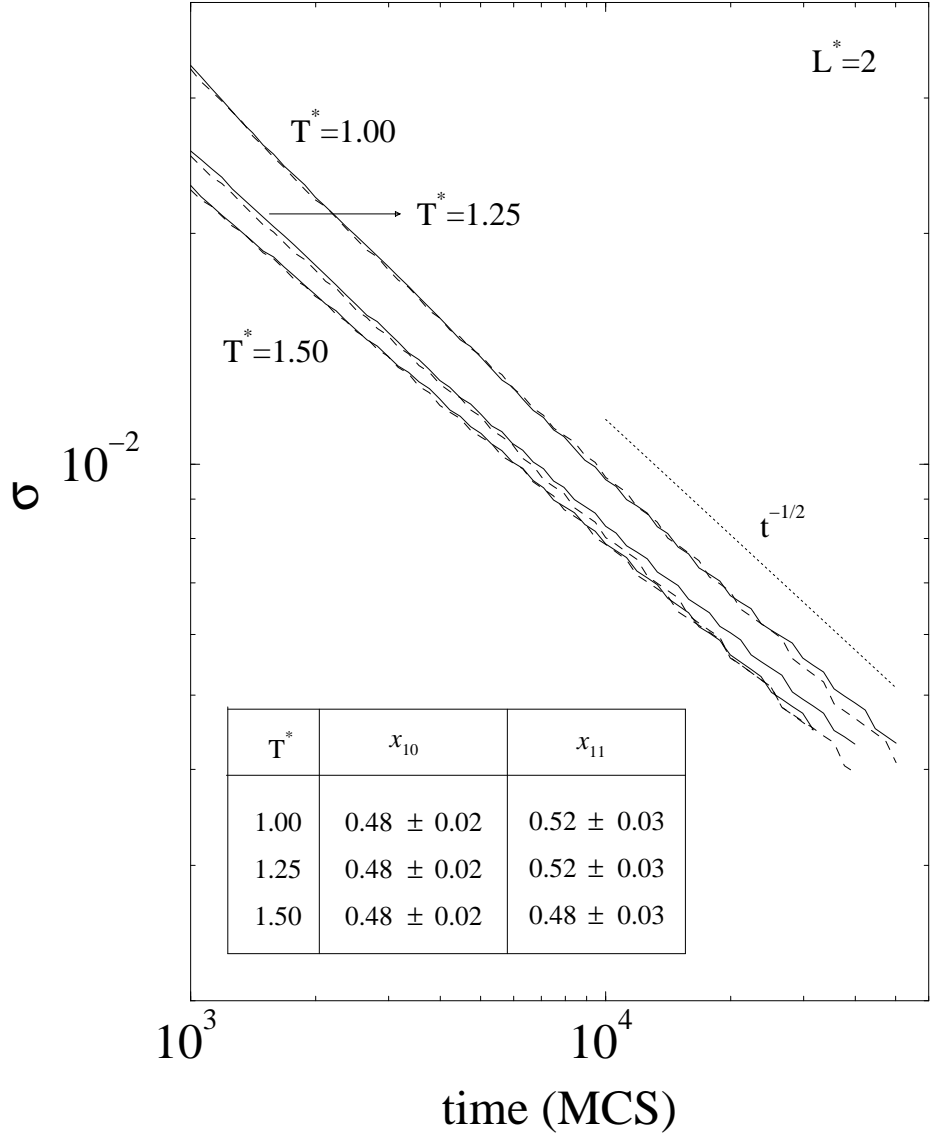}
\caption{Evolution of the structure factor width $\sigma(t)$, for a system 
with $L^*=2$ and $\ell=600$, along the two directions $(10)$ (continuous lines) 
and $(11)$ (discontinuous lines). Data corresponds to three different 
temperatures, as indicated. The dotted straight line indicates the Allen-Cahn
 behavior $\sigma (t) \sim t^{-1/2}$. The
inset shows the fitted power-law exponents to the last 25000 MCS.}
\label{FIG8}
\end{figure}

We now present the data obtained from the study of the structure factor.  In
Fig.  \ref{FIG8} we show the time evolution of the width of the superstructure 
peak, $\sigma(t)$, in the two relevant directions and for the same three 
temperatures and model parameters as in
Figures \ref{FIG4} and \ref{FIG5}.  The long-time behavior is, for the three
temperatures, algebraic ($\sigma \sim t^{-x}$) and characterized by an exponent
$x=0.50 \pm 0.05$.  The inset contains the values of the growth exponent 
obtained
by fitting the algebraic law $\sigma(t) \sim t^{-x}$ to the data at each
temperature.

\begin{figure}
\psboxto(0.8\textwidth;0cm){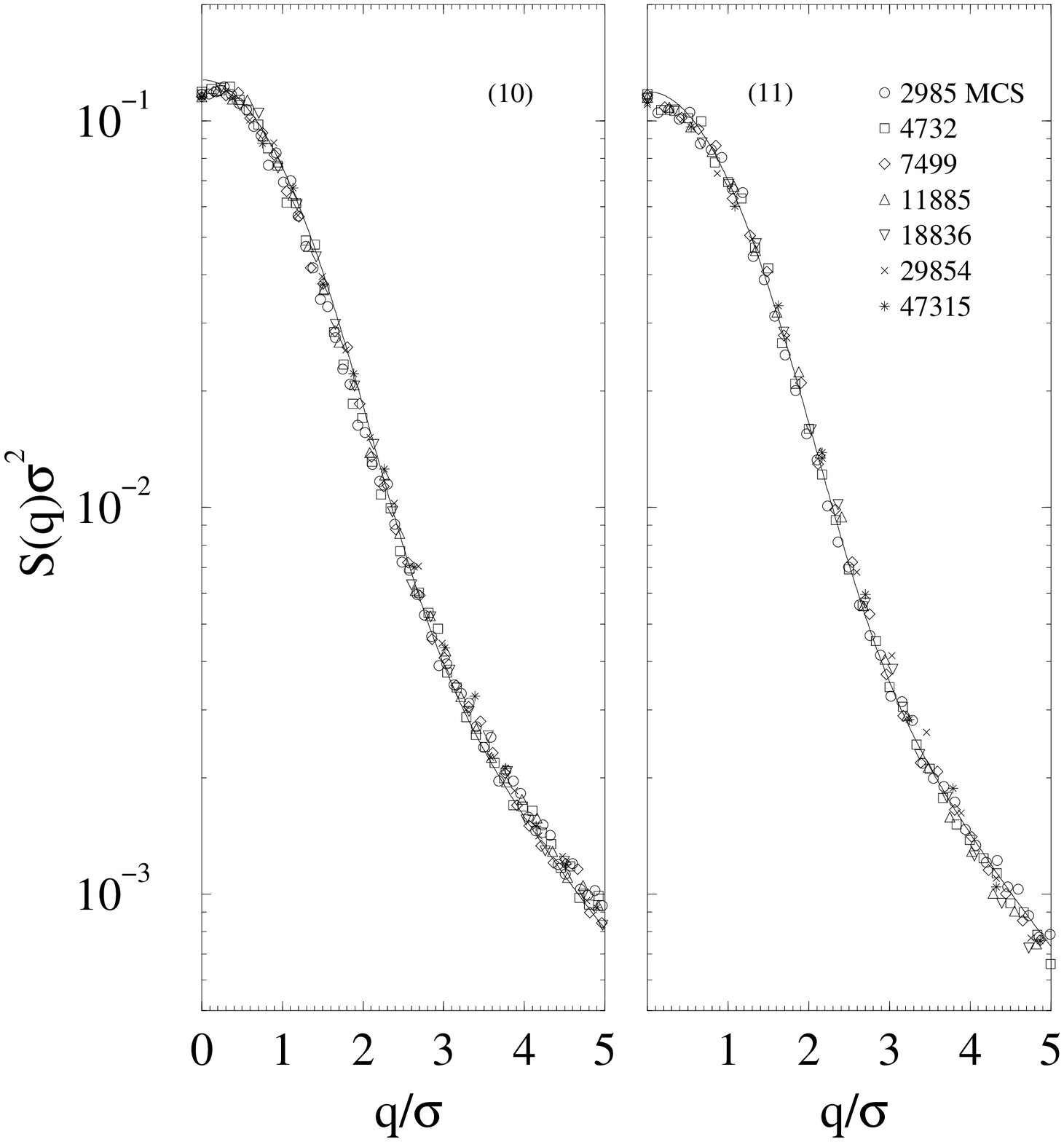}
\caption{Scaling of the structure factor profiles for a system with $L^*=2$,
$\ell = 600$ and $T^*=1.0$, along the two directions $(10)$ and 
$(11)$. The continuous line corresponds to the universal scaling function 
proposed by Ohta {\it et al.} (Ref. 23).}
\label{FIG9}
\end{figure}

Moreover we have studied the existence of dynamical scaling. As an example,
in Figure \ref{FIG9} 
we show the structure factor profiles at $T^*=1.0$ scaled with the 
corresponding $\sigma(t)$ along the two directions. The overlap of the data is 
very satisfactory. Moreover the shape of the scaling function can be compared
with the universal scaling function corresponding to a non-conserved
order parameter system \cite{Ohta82}, which is also plotted with a continuous 
line. The agreement is excellent for large and intermediate values of $q/\sigma$. 
For small 
values $q/\sigma < 1$ the behavior is slightly different in both 
directions.
This may be associated to the existence of anisotropic correlations in the
domain shape. This will be analyzed in detail in a future work.

\section{Summary and Conclusions}

The results presented in the previous section show that during the asymptotic 
regime of vacancy-assisted domain-growth $\Delta E(t)^{-1} \sim m(t) 
\sim\sigma(t)^{-1}$. In spite of some details observed for small values of q, 
one may 
conclude that dynamical scaling holds and that the long-time behavior can be 
characterized by an unique length which evolves with time
according to $R(t) \sim t^{1/2}$.  This asymptotic regime is preceded by a
long transient that depends on both the temperature and the asymmetry 
character of
the alloy.  In the ABV model the asymmetry is controlled by the parameter $L^*$
which in turn determines the energy barrier $\epsilon^* = 4(|L^*|-1)$ that the
vacancy has to climb in its migration from the bulk to the interphases.  The
characteristics of the transient regime are determined by the ratio 
$\epsilon^*/T^*$.

This behavior is definitively different from the one found in symmetric binary
alloys with $L^*=0$.  The origin of this difference lies in the fact that for
symmetric binary alloys the interaction between the vacancies and the interphases
is attractive.  The first consequence of this is to speed up the global 
process.
The growth exponent was found to be temperature dependent \cite{Frontera94} 
and clearly larger
that $1/2$ at low and moderate temperatures.  Only at $T^* \rightarrow T_{c}^*$
the Allen-Cahn exponent is recovered.

The fast growth predicted in the symmetric case has
never  been   found experimentally.  The  authors   \cite{Vives92,Vives93}   
argued   that  the
temperatures  at which  experiments are usually  performed are too high.  In the
light of the present  results it seems more reliable  that it has to do with the
intrinsic asymmetric character of real alloys. Actually the asymmetry parameter 
$L^*$  can be estimated from  
ab-initio calculations. We have used the  WIEN97 code \cite{wien} to evaluate 
$L^*$ for the  $\beta$-CuZn alloy and have obtained $L^* \cong 3$. \cite{marcel}

\acknowledgements 

We acknowledge financial support from the Comisi\'on Interministerial de Ciencia
y Tecnolog\'{\i}a  (CICyT, project number MAT98-0315).  One of us (M.P.)  thanks
K.  Schwarz,  P.  Blaha  and P.  Mohn for  helpful  discussions  and for  their
hospitality  during his stay at the Vienna  University of  Technology  where the
computation of the model  parameters of $\beta$-CuZn  was performed.  M.P.  also
acknowledges financial support from the Comissionat per a Universitats i Recerca
(Generalitat de Catalunya).


\begin{table}
\caption{Absolute bond energies and excess energies with respect to the $A-B$
bond for the ABV model as  a function of the  parameters $J$, $K$, and
$L$.}
\begin{tabular}{ccc} Bond & Energy & Excess energy \\ \hline $A-B$ & $-J+K$ & 0
\\ $A-A$ & $J+K+2L$ & $2J+2L$ \\ $B-B$ & $J+K-2L$ & $2J-2L$ \\ $A-V$ &
$0$ & $J-K$ \\ $B-V$ & $0$ & $J-K$ \\ $V-V$ & $0$ & $J-K$
\end{tabular} 
\label{TABLE1}
\end{table}

\newpage

\begin{references}


\bibitem{Gunton83} J.D.Gunton, M.San   Miguel and P.S.Sahni,   in {\sl
Phase   Transitions and Critical  Phenomena},   edited  by C.Domb  and
J.L.Lebowitz (Academic Press., New York, 1983), Vol. 8.

\bibitem{Komura88} {\sl   Dynamics of Ordering  Processes in Condensed
Matter}, edited by S.Komura and  H.Furukawa (Plenum Press., New  York,
1988).

\bibitem{Bray94}A. J. Bray, Adv. Phys. {\bf 43}, 357 (1994). 


\bibitem{Phani80} M.K.Phani,  J.L.Lebowitz, M.H.Kalos  and  O.Penrose,
Phys. Rev.  Lett. {\bf 45}, 366 (1980).

\bibitem{Sahni81}   P.S.Sahni,   G.Dee,     J.D.Gunton,     M.Phani,
J.L.Lebowitz and M.Kalos, Phys. Rev. B {\bf 24},410 (1981).

\bibitem{Fogedby88} H.C.Fogedby and  O.G.Mouritsen, Phys. Rev.  B {\bf
37}, 5962 (1988) and references therein.


\bibitem{Allen79} S.M.Allen and J.W.Cahn, Acta  Metall. {\bf 27}, 1085
(1979).

\bibitem{Mouritsen90} O.G.Mouritsen, Int. J.   Mod.  Phys.  B {\bf 4},
1925 (1990).


\bibitem{Frontera94} C.Frontera, E.Vives and  A.Planes,  Z.  Phys.   B
{\bf 96}, 79 (1994).

\bibitem{Vives92} E.Vives and   A.Planes, Phys.  Rev.  Lett.  {\bf 68}, 812
(1992).

\bibitem{Vives93} E.Vives and A.Planes, Phys.  Rev.  B {\bf 47}, 2557 (1993).


\bibitem{Frontera93} C.Frontera, E.Vives and  A.Planes, Phys.  Rev.  B
{\bf 48} 9321, (1993).


\bibitem{Neumann76} P.Neumann, Y.A.Chang and C.M.Lee, Acta Metall {\bf24}, 593 
(1976); J.Mayer,       C.Els\"asser  and    M.F\"ahnle,
Phys. Stat. Sol.(b) {\bf 191}, 283 (1995) ; and references therein.

\bibitem{Frontera97}  C.Frontera, Thesis,    Universitat de  Barcelona
(1997).


\bibitem{Marcel97}   M.Porta,    C.Frontera,   E.Vives and T.Cast\'an,
Phys. Rev. B {\bf 56}, 5261 (1997).

\bibitem{Srolovitz87} D.J.Srolovitz and G.N.Hassold, Phys. Rev. B {\bf
35}, 6902 (1987).

\bibitem{Mouritsen89} O.G.Mouritsen and  P.J.Shah, Phys.  Rev.  B {\bf
40}, 11445 (1989).

\bibitem{Shah90} P.J.Shah and  O.G.Mouritsen, Phys.  Rev.  B {\bf  41},
7003 (1990).

\bibitem{Binder88}   K.Binder  and  D.W.Heermann,  {\sl    Monte Carlo
simulation in Statistical Physics}, Springer-Verlag (Berlin 1988).


\bibitem{Fultz87} B. Fultz, J.Chem. Phys. {\bf 87}, 1604 (1987); 
H.C.Kang and W.H. Weinberg, J. Chem. Phys {\bf 90}, 2824 (1989).


\bibitem{cvalloys} R.E.Smallman, {\sl Modern Physical Metallurgy} (Butterworths, 
London 1970). 

\bibitem{ordpar} In two dimensions,  $m(t) \propto R(t)$, as shown  in
K. Binder, J. Comp. Phys. {\bf 59}, 1 (1985).

\bibitem{Ohta82} T.Ohta, D.Jasnow and K.Kawasaki, Phys. Rev. Lett. {\bf 49}, 
1223 (1982).


\bibitem{wien}  P.  Blaha, K.   Schwarz,  and J. Luitz, {\bf  WIEN97},
Vienna University   of Technology 1997.   (Improved and   updated Unix
version of the original copyrighted WIEN-code,  which was published by
P.    Blaha,  K.   Schwarz, P.  Sorantin,    and  S.  B.   Trickey, in
Comput. Phys. Commun. 59, 399 (1990)).


\bibitem{marcel} 
We use the ${\bf WIEN}$ package  which is based on the full potential linearized
augmented plane wave method.  The parameter $L^*$ is estimated by comparing the
energy changes due to n.n. vacancy jumps in different partially ordered 
neighborhoods, with the corresponding energies given by an ABV model defined 
on a BCC lattice with interactions up to n.n.n. 




\end{references}
\end{document}